\documentclass[a4paper,fleqn,usenatbib]{mnras}

\usepackage[T1]{fontenc}
\usepackage{ae,aecompl}

\usepackage{graphicx}	
\usepackage{amsmath}	
\usepackage{amssymb}	

\usepackage[dvipsnames,usenames,table]{xcolor}

\title[A fast bar in NGC~4264]{Evidence of a  fast bar in the weakly-interacting galaxy NGC~4264 with MUSE}

\author[V. Cuomo et al.]{V. Cuomo,$^{1}$\thanks{E-mail: virginia.cuomo@phd.unipd.it}
E. M. Corsini,$^{1,2}$
J. A. L. Aguerri,$^{3,4}$
V. P. Debattista,$^{5}$
L. Coccato,$^{6}$
\newauthor{L. Costantin,$^{1,7}$
E. Dalla Bont\`a,$^{1,2}$
E. Iodice,$^{8}$}
J. M\'endez-Abreu,$^{3,4}$
L. Morelli,$^{9}$
\newauthor{I. Pagotto,$^{1}$
and A. Pizzella$^{1,2}$}
\\
$^{1}$Dipartimento di Fisica e Astronomia "G. Galilei", Universit\`a di Padova, vicolo dell'Osservatorio 3, I-35122 Padova, Italy\\
$^{2}$INAF - Osservatorio Astronomico di Padova, vicolo dell'Osservatorio 2, I-35122 Padova, Italy\\
$^{3}$Departamento de Astrof\'isica, Universidad de La Laguna, Avenida Astrof\'isico Francisco S\'anchez s/n, 38206 La Laguna, Tenerife, Spain\\
$^{4}$Instituto de Astrof\'isica de Canarias, calle V\'ia L\'actea s/n, 38205 La Laguna, Tenerife, Spain\\
$^{5}$Jeremiah Horrocks Institute, University of Central Lancashire, PR1 2HE Preston, UK\\
$^{6}$European Southern Observatory, Karl-Schwarzschild-Strasse 2, D-85748 Garching, Germany\\
$^{7}$INAF -  Osservatorio Astronomico di Brera, via Brera, 28, I-20159 Milano, Italy\\
$^{8}$INAF - Osservatorio Astronomico di Capodimonte, via Moiariello 16, I-80131 Napoli, Italy\\
$^{9}$Instituto de Astronom\'ia y Ciencias Planetarias, Universidad de Atacama, Avenida Copayapu 485, Copiap\'o, Chile
}

\date{Accepted 2019 June 21. Received 2019 June 10; in original form 2019 February 20}

\pubyear{2019}

\begin{document}
\label{firstpage}
\pagerange{\pageref{firstpage}--\pageref{lastpage}}
\maketitle

\begin{abstract}
We present surface photometry and stellar kinematics of NGC~4264, a barred lenticular galaxy in the region of the Virgo Cluster undergoing a tidal interaction with one of its neighbours, NGC~4261. We measured the bar radius ($a_{\rm bar}=3.2 \pm 0.5$ kpc) and strength ($S_{\rm bar}=0.31 \pm 0.04$) of NGC~4264 from Sloan Digital Sky Survey imaging and its bar pattern speed ($\Omega_{\rm bar}=71\pm 4$ km s$^{-1}$ kpc$^{-1}$) using the Tremaine-Weinberg method with stellar-absorption integral-field spectroscopy performed with the Multi Unit Spectroscopic Explorer at the Very Large Telescope. We derived the circular velocity ($V_{\rm circ}= 189\pm  10$ km s$^{-1}$) by correcting the stellar streaming velocity for asymmetric drift and calculated the corotation radius ($R_{\rm cor}=2.8 \pm 0.2$ kpc) from the bar pattern speed. Finally, we estimated the bar rotation rate ($R_{\rm cor}/a_{\rm bar}=0.88\pm 0.23$). We find that NGC~4264 hosts a strong and large bar extending out to the corotation radius. This means that the bar is rotating as fast as it can like nearly all the other bars measured so far even when the systematic error due to the uncertainty on the disc position angle is taken into account. The accurate measurement of the bar rotation rate allows us to infer that the formation of the bar of NGC~4264 was due to self-generated internal processes and not triggered by the ongoing interaction.
\end{abstract}



\section{Introduction}

Although unbarred galaxies were defined as ``normal'' by \citet{Hubble1926} and ``ordinary'' by \citet{deVaucouleurs1959} in their morphological classifications, they do not constitute the majority of disc galaxies in the local universe. 
On the contrary, $\sim70$ per cent of them host a bar or have weaker non-axisymmetric features of a similar kind.  Indeed, a large number of galaxies which appeared unbarred in the blue photographic plates used in the early classifications turned out to be barred when imaged by digital detectors in red and near-infrared passbands \citep[e.g.,][]{Knapen2000, Aguerri2009, Nair2010, Buta2015}. 

The observed motions within the bar are consistent with most of the stars streaming along highly-elongated regular orbits aligned with the bar major axis \citep{Contopoulos1980, Manos2011}. According to theoretical predictions, a fraction of stochastic orbits is also present \citep{Martinet1990, Patsis2014}. The bar pattern tumbles about the rotation axis of the galaxy normal to the disc plane. Gas flow patterns seem to be well understood too, as they drive the formation of the offset dust lanes observed at the edges of many bars \citep{Athanassoula1992, Kim2012}. Both the gaseous and stellar distributions are expected to evolve on a time-scale of many bar rotation periods because the bar is responsible for a substantial redistribution of mass and angular momentum in the disc. Therefore, both the morphology and dynamics of a barred galaxy depends on the bar pattern speed $\Omega_{\rm bar}$ which is the angular speed of rotation of the bar as viewed from an inertial frame \citep{Athanassoula2003, Combes2011}. 

To relate the predictions of theoretical works and results of numerical experiments to real galaxies, the bar pattern speed is parametrised with the bar rotation rate ${\cal{R}}\equiv R_{\rm cor}/a_{\rm bar}$. This is the distance-independent ratio between the corotation radius $R_{\rm cor}$ and the bar semi-major axis $a_{\rm bar}$, which corresponds to the bar radius. When the rotation curve is flat, the corotation radius is derived from   the bar pattern speed as $R_{\rm cor} = V_{\rm circ}/\Omega_{\rm bar}$, where $V_{\rm circ}$ is the circular velocity.
As far as the value of ${\cal{R}}$ is concerned, dynamical arguments show that if ${\cal{R}} < 1.0$ the stellar orbits are elongated perpendicular to the major axis of the bar and it dissolves. Bars with $1.0 \leq {\cal{R}} \leq 1.4$ end close to corotation and rotate as fast as they can, whereas bars with ${\cal{R}} > 1.4$ fall short of corotation and are termed slow. The dividing value at 1.4 between long/fast and short/slow bars is given by consensus \citep{Athanassoula1992, Debattista2000} and it does not imply a specific value of the pattern speed.

Both analytical work \citep{Weinberg1985} and numerical simulations \citep[e.g.,][]{Little1991, Debattista1998, ONeill2003, VillaVargas2010, Athanassoula2013} show that the bar pattern speed decreases with time as a consequence of the angular momentum exchange within the galaxy and the dynamical friction exerted on the bar by the dark matter (DM) halo. In both cases, a massive and centrally-concentrated DM halo causes a slow down of the bar because there is more mass ready to absorb angular momentum near the resonances and the dynamical friction is more efficient (see also \citealt{Athanassoula2014} and \citealt{Sellwood2014} for a further discussion). This allowed \citet{Debattista2000} to put tight constraints on the DM distribution in barred galaxies and argue that galaxies hosting fast bars should be embedded in DM halos with a low central density, such as those required for maximum discs. This makes the measurement of the rotation rate of bars highly desirable not only to investigate the secular evolution of barred galaxies but also to test whether the measured DM distribution matches that predicted by cosmological simulations with cold DM \citep{Navarro1996, Moore1998, Zasov2017}.

The most straightforward way to derive the bar pattern speed is the technique developed by \citet[hereafter TW]{Tremaine1984}, which measures the average position and velocity of a tracer population that obeys the continuity equation along different cuts crossing the bar and parallel to the disc major axis. This method is best suited to the analysis of the distribution and kinematics of the old stellar component in the absence of significant star formation and patchy dust obscuration \citep{Gerssen2007}.  TW observations provide an upper limit on the intrinsic ellipticity of discs in agreement with photometric constraints, although the scatter of $\cal{R}$ produced by the disc non-axisymmetry may be significant \citep{Debattista2003}.

Early applications of the TW method based on long-slit spectroscopy were challenging in terms of both integration times and kinematical analysis and therefore focused on early-type barred galaxies \citep[see][for a review]{Corsini2011}. Over a dozen galaxies, including a double-barred galaxy \citep{Corsini2003}, were measured with a typical uncertainty of $\sim30$ per cent mostly due to errors in identifying the position of the galaxy centre and in measuring the galaxy systemic velocity, low signal-to-noise ratio ($S/N$) of the spectra, limited number of the slits and their misalignment with respect to the disc major axis. The advent of integral-field spectroscopy on wide field of views promises to overcome these problems and lead to more efficient and precise TW measurements \citep[but see][for a first application]{Debattista2004}. Indeed, the centring errors in both the position of the galaxy centre and in measuring the galaxy systemic velocity are minimised by the unambiguous determination of the common reference frame for the distribution and velocity field of the stars, the $S/N$ of the spectra can be increased by rebinning adjacent spaxels, and the number and orientation of the pseudoslits can be optimized during the analysis.

\citet{Aguerri2015} measured the bar pattern speed of 15 galaxies on the stellar velocity maps provided by CALIFA integral-field spectroscopic survey \citep{Sanchez2012}. More recently, \cite{Guo2019} obtained the bar pattern speed for another 51 galaxies\footnote{The paper lists 53 objects but the galaxies 8274-6101 and 8603-12701 are duplications of 8256-6101 and 8588-3701, respectively.} using the integral-field spectroscopic data from MaNGA project \citep{Bundy2015}. Neither of them found significant trends between $\cal{R}$ and morphological type although the two samples cover the entire sequence of barred galaxies from SB0s to SBds. The fast bar solution can not be ruled out for any galaxy, in agreement with results from indirect measurements of the bar pattern speed. However, the typical uncertainty of the bar pattern speeds of the CALIFA and MaNGA galaxies is $\sim 30$ and $\sim 50$ per cent, respectively, because of the limited spatial sampling of the spectroscopic data which restricted the TW analysis to only three to five pseudoslits.

In this paper we derive the bar pattern speed of the lenticular galaxy NGC~4264 from integral-field spectroscopy performed with the Multi Unit Spectroscopic Explorer (MUSE) at the Very Large Telescope (VLT). With this pilot study we aim to show that integral-field spectroscopic data with high spatial sampling are mandatory to substantially reduce the uncertainty on the bar pattern speeds measured with the TW method and properly compare the observed bar rotation rates with theoretical predictions and results of numerical simulations. We structure the paper as follows. We present the general properties of NGC~4264 in Sec.~\ref{sec:properties}. We show the broad-band imaging in Sec.~\ref{sec:imaging}  and the integral-field spectroscopy in Sec.~\ref{sec:spectroscopy}. We derive the bar properties in Sec.~\ref{sec:bar} and discuss our findings in Sec.~\ref{sec:conclusions}. 

\section{Main properties of NGC~4264}
\label{sec:properties}

NGC~4264 is an early-type disc galaxy which was classified as SB0 by \cite{UGC}, as SB0$^+$(rs) by \citet[hereafter RC3]{RC3}, and as SBa by \citet{Kim2014}. 

\begin{figure}
    \centering
    \includegraphics[scale=0.5]{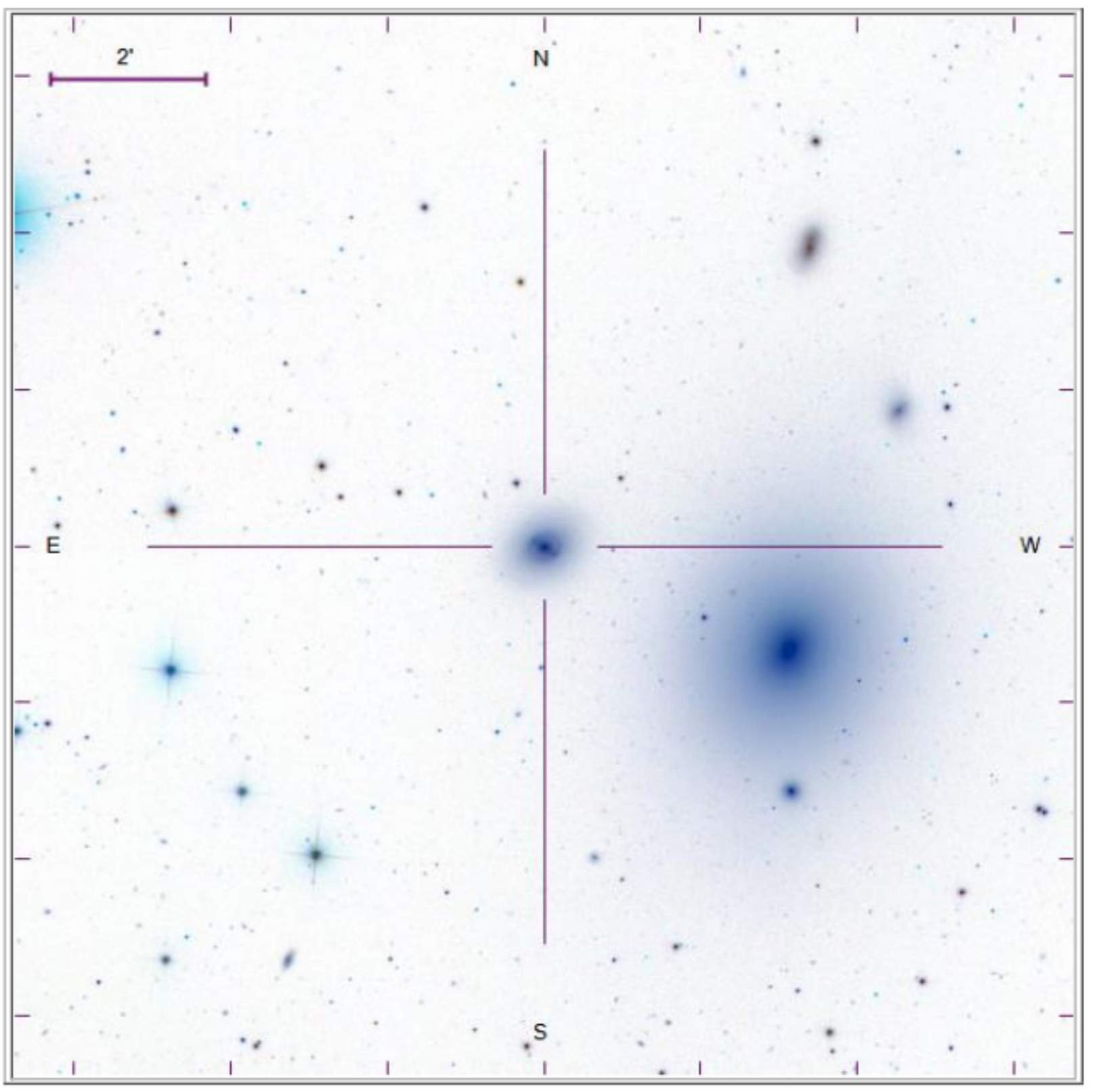}
    \caption{SDSS $i$-band image of NGC~4264 and NGC~4261. The size and orientation of the FOV are given and a cross marks the centre of NGC~4264.}
    \label{fig:sdss_image}
\end{figure}

It is characterised by an apparent magnitude $B_{T}=13.70$ mag (RC3), which corresponds to a total corrected absolute magnitude $M^0_{B_{\rm T}}=-19.27$ mag, obtained adopting a distance $D = 39.2$ Mpc from the radial velocity with respect to the cosmic microwave background reference frame $V_{\rm CMB} = 2864\pm25$ km s$^{-1}$ \citep{Fixen1996} and assuming $H_0=73$ km s$^{-1}$ Mpc$^{-1}$. Nevertheless, the galaxy was classified as a possible member of the Virgo Cluster \citep{Kim2014} and it belongs to the rich galaxy group around the early-type galaxy NGC~4261 \citep{Garcia1993, Kourkchi2017}. According to \cite{Schmitt2001}, NGC~4264 possibly forms an interacting couple with NGC~4261 (Fig.~\ref{fig:sdss_image}) which lies at a projected distance of 3.5 arcmin (30 kpc). But, they also pointed out that this interaction is not a necessary condition to trigger the nuclear activity of NGC~4261. 

\citet{Cappellari2013, Cappellari2013bis} constructed a Jeans axisymmetric dynamical model to constrain the orbital structure of the stars and the DM content within the half-light radius $R_{\rm e}$ by matching the galaxy surface brightness and stellar kinematics available from the ATLAS$^{3 \rm D}$ project. They found a stellar mass-to-light ratio $\log (M/L_r)_{\rm stars}=0.445$ scaled to the adopted distance and DM fraction $f{\rm (DM)}=0.31$ within a sphere of radius $R_{\rm e}$ in the $r$-band of the Sloan Digital Sky Survey (SDSS). 

\section{Broad-band imaging}
\label{sec:imaging} 

\subsection{Image acquisition and reduction}
\label{sec:image_reduction}

We retrieved the $g$- and $i$-band images of NGC~4264 from the Data Archive Server (DAS) of the Data Release 12 of the SDSS  (SDSS-DR12, \citealt{Alam2015}). The images were already bias-subtracted, flatfield-corrected, sky-subtracted, and flux-calibrated according to the associated calibration information stored in the DAS.

We trimmed the images selecting a field of view (FOV) of $800\times800$ pixel ($5.3\times5.3$ arcmin$^2$) centred on the galaxy (Fig.~\ref{fig:sdss_image}). To estimate the goodness of the SDSS sky subtraction, we fitted elliptical isophotes with the {\sc ellipse} task in {\sc iraf} \citep{Jedrzejewski1987} to measure the radial profile of the surface brightness at large distance from the galaxy centre. We masked foreground stars, nearby and background galaxies, residual cosmic rays, and bad pixels before fitting the isophotes. As a first step, we allowed the centre, ellipticity, and position angle of ellipses to vary. Then, we adopted the centre of the inner ellipses ($R<2$ arcsec) and the ellipticity and position angle of the outer ones ($R>180$ arcsec). The radial profile of the background surface brightness shows a remarkable gradient in both $g$- and $i$-band images due to the residual light contribution of the bright nearby galaxy NGC~4261.

Since NGC~4261 is not fully targeted in the FOV of the NGC~4264 images, we retrieved also the $g$- and $i$-band images of NGC~4261 from SDSS-DR12. Then we performed a photometric decomposition of NGC~4261 by using the Galaxy Surface Photometry 2-Dimensional Decomposition algorithm ({\sc gasp2d}, \citealt{MendezAbreu2008, MendezAbreu2014, deLorenzoCaceres2019}). We modelled the surface brightness distribution of the galaxy with a S\'ersic law following the prescriptions given in Sec.~\ref{sec:gasp2d}. The model image of NGC~4261 was convolved with a circular Moffat point spread function \citep[PSF,][]{Moffat1969} with the shape parameters measured directly from the field stars of the NGC~4264 image and then it was subtracted from the image of NGC~4264. We conducted this analysis for
both the $g$- and $i$-band images.

Finally, we repeated the ellipse fitting of the isophotes with constant centre, ellipticity, and position angle on the corrected images of NGC~4264. We found a constant surface brightness for $R\sim140$ arcsec, which we adopted as the residual sky level to be subtracted from the image. We measured the standard deviation of the image background after the residual sky subtraction in regions free of sources at the edges of the FOV (Fig.~\ref{fig:sdss_image}) using the {\sc iraf} task {\sc imexamine}. We found $\sigma_{{\rm sky},g} = 0.07$ and $\sigma_{{\rm sky},i} = 0.04$ mag arcsec$^{-2}$, while the sky surface brightness was $\mu_{{\rm sky},g} = 24.80$ and $\mu_{{\rm sky},i} = 23.69$ mag arcsec$^{-2}$.

\subsection{Isophotal analysis}
\label{sec:isophotal_analysis}

We performed the isophotal analysis of the sky-subtracted images of NGC~4264 in both $g$- and $i$-bands using {\sc ellipse}. We fitted the galaxy isophotes with ellipses fixing the centre coordinates after checking they do not vary within the uncertainties. The resulting radial profiles of the azimuthally-averaged surface brightness $\mu$, position angle PA, and ellipticity $\epsilon$ in the $i$-band are shown in Fig.~\ref{fig:gasp2d}.

We measured no colour variation over the observed radial range ($\mu_g-\mu_i=1.148\pm0.006$ mag arcsec$^{-2}$) and the same radial profiles of PA and $\epsilon$ in both passbands. The PA decreases from ${\rm PA}\sim95\degr$ to ${\rm PA}\sim60\degr$ in the inner 10 arcsec, where $\epsilon$ peaks to $\epsilon\sim0.37$. The PA steadily rises outwards to ${\rm PA}\sim110\degr$ at $R\sim16$ arcsec, while $\epsilon$ falls to $\epsilon\sim0.15$ at $R\sim13$ arcsec and it increases to $\epsilon\sim0.19$ at $R\sim16$ arcsec. The PA shows a constant behaviour of ${\rm PA}\sim60\degr$ around $R \sim9$~arcsec, corresponding to the region of the peak in $\epsilon$. 

The presence of a local maximum in the $\epsilon$ radial profile which corresponds to a nearly constant PA is a typical feature of barred galaxies \citep[e.g.,][]{Wozniak1995, Aguerri2009} and is due to the shape and orientation of the stellar orbits of the bar \citep[e.g.,][]{Contopoulos1989,  Athanassoula1992}. Further out, the PA of NGC~4264 rises to ${\rm PA}\sim120\degr$ at the farthest measured radius, while $\epsilon$ remains constant. 
To quantify such an isophotal twist, we derived the mean PA of the galaxy isophotes in two different radial ranges corresponding to the inner ($18<R_{\rm in} <23$ arcsec) and outer ($27<R_{\rm out} <41$ arcsec) portion of the disc, respectively. We fixed the lower limit of the inner radial range just outside the bar-dominated region and the upper limit of the outer range at the farthest observed radius. We defined the extension of the radial ranges by fitting the PA measurements with a straight line and considered all the radii where the line slope was consistent with being zero within the associated root mean square error. The two regions have the same $\epsilon$ ($\epsilon_{\rm in}=0.199\pm 0.008$, $\epsilon_{\rm out} = 0.20\pm 0.02$) and therefore the discs have the same inclination ($i_{\rm in} = 36\fdg7 \pm 0\fdg7$, $i_{\rm out}= 37^\circ \pm 1^\circ$ for an infinitesimally thin disc) but are characterised by a significantly different PA ($\rm PA_{\rm in} = 114\fdg0\pm 1\fdg2$, $\rm PA_{\rm out}= 122\fdg8\pm 2\fdg4$). 

Previous measurements of the PA and $\epsilon$ of NGC~4264 were obtained by \cite{Krajnovic2011} by fitting the galaxy surface-brightness distribution from the SDSS $r$-band image using all the available radial range and with no distinction between the inner and outer region of the disc. They found ${\rm PA}=119\fdg8 \pm 5\fdg5$ which is in between and consistent within the errors with our two estimates, and $\epsilon=0.19\pm 0.01$ which fully agrees with our findings.

\subsection{Photometric decomposition}
\label{sec:gasp2d}

We derived the structural parameters of NGC~4264 by applying the {\sc gasp2d} algorithm to the sky-subtracted SDSS $i$-band image of the galaxy. We modelled the galaxy surface brightness in each image pixel to be the sum of the light contribution of a S\'ersic bulge \citep{Sersic1968}, a double-exponential disc \citep{MendezAbreu2017}, and a Ferrers bar \citep{Aguerri2009}. We assumed that their isophotes are elliptical and centred on the galaxy centre ($x_0,y_0$) with constant values of position angles PA$_{\rm bulge}$, PA$_{\rm disc}$, PA$_{\rm bar}$ and axial ratios $q_{\rm bulge}$, $q_{\rm disc}$, and $q_{\rm bar}$, respectively. We did not account for other luminous components, such as rings or spiral arms.

The best-fitting values of the structural parameters of the bulge, disc, and bar are returned by {\sc gasp2d} by performing a $\chi^2$ minimization. We weighted the surface brightness of the image pixels according to the variance of the total observed photon counts due to the contribution of both galaxy and sky, which we calculated by taking into account for photon noise, gain and read-out noise of the detector. We adopted the same mask image built for the isophotal analysis and excluded the masked pixels from the fit. We handled the seeing effects by convolving the model image with a circular Moffat PSF with the shape parameters (${\rm FWHM}=1.18$ arcsec, $\beta=2.99$)  measured directly from the stars in the image. We hold fixed $q_{\rm disc}=0.796$ because of the constant $\epsilon$ measured from the isophotal analysis and $r_{\rm break} = 24.3$ arcsec (= 61 pixel) as the end of the inner disc from a visual inspection of the surface-brightness radial profile to allow to find the remaining parameters. We adopted a double exponential law for the disc after checking that the residuals in the outer regions have a median value consistent with 0 mag arcsec$^{-2}$ whereas they systematically rises from 0 to 0.4 mag arcsec$^{-2}$ if a single exponential is adopted.

\begin{figure*}
\includegraphics[scale=0.6, angle=90]{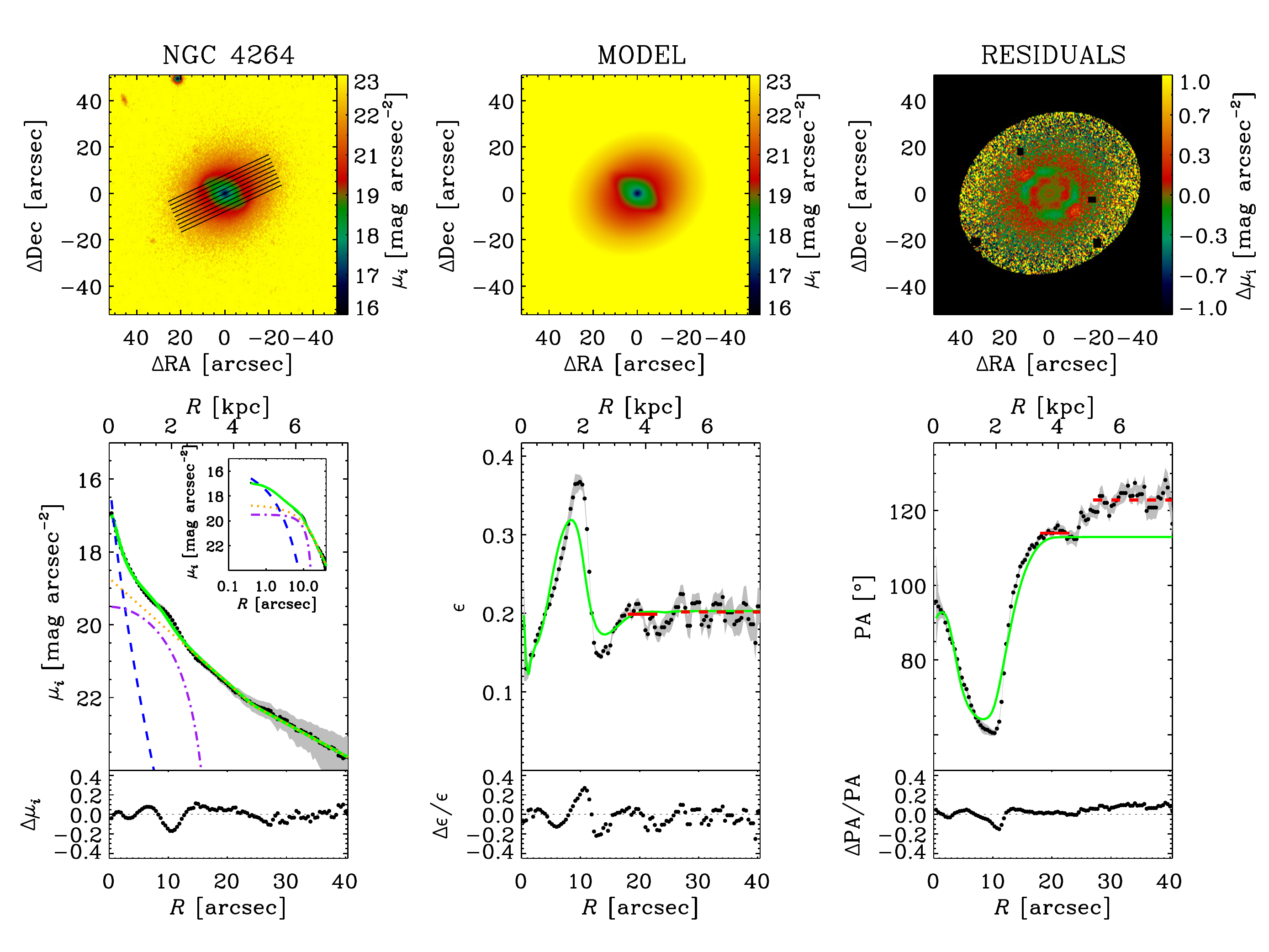}
\caption{Two-dimensional photometric decomposition of the SDSS $i$-band image of NGC~4264 as obtained from {\sc gasp2d}. The upper panels (from left to right) show the map of the observed, modelled, and residual (observed$-$modelled) surface brightness distributions. The FOV is oriented with North up and East left. The diagonal black lines overplotted to the observed image show the location of the pseudoslits adopted to derive the bar pattern speed. The black areas in the residual image correspond to pixels excluded from the fit. The lower panels (from left to right) show the radial profile of surface brightness, PA, and $\epsilon$ from the isophotal analysis of the observed (black dots with gray error bars) and seeing-convolved modelled image (green solid line) and their corresponding difference. The surface brightness radial profiles of the best-fitting bulge (blue dashed line), bar (magenta dash-dotted line), and disc (orange dotted line) are also shown in both linear and logarithmic scale for the semi-major axis distance to the centre of the galaxy. The horizontal red solid and dashed segments in the PA and $\epsilon$ panels give the mean values measured for the inner and outer portions of the disc, respectively, and mark the extension of the radial ranges which we adopted to calculate them.}
\label{fig:gasp2d}
\end{figure*}

Figure~\ref{fig:gasp2d} shows the SDSS $i$-band image, {\sc gasp2d} best-fitting image, and residual image of NGC~4264. The values of its best-fitting structural parameters and corresponding errors are reported in Table~\ref{tab:structural_parameters}. 

We estimated the errors on the best-fitting structural parameters of NGC~4264 by analysing the images of a sample of mock galaxies generated by \citet{MendezAbreu2017} with Monte Carlo simulations and mimicking the instrumental setup of the available SDSS image. They assumed their mock galaxies to be at a distance of 67 Mpc which is the median value of their sample, after checking that our galaxy is in the same distance range. Moreover, we analysed the barred galaxies with total apparent magnitude in the range $12 \leq m_i \leq 13$ mag to match the characteristics of NGC~4264. For the bulge ($I_{\rm e}$, $r_{\rm e}$, $n$), disc ($I_{0,{\rm disc}}$, $h_{\rm in}$, $h_{\rm out}$), and bar surface-brightness parameters ($I_{0,{\rm bar}}$, $a_{\rm bar}$), we adopted the mean and standard deviation of the relative errors of the mock galaxies as the systematic and statistical errors of the observed galaxies, respectively. For $q_{\rm bulge}$, $q_{\rm bar}$, PA$_{\rm bulge}$, PA$_{\rm disc}$, and PA$_{\rm bar}$ we adopted the mean and standard deviation of the absolute errors of the mock galaxies as the systematic and statistical errors $\sigma_{\rm syst}$ and $\sigma_{\rm stat}$ of the observed galaxies, respectively. We computed the errors as $\sigma^2 = \sigma^2_{\rm stat} + \sigma^2_{\rm syst}$, with the systematic errors negligible compared to the statistical ones.

NGC~4264 hosts a small and nearly exponential bulge and a large anti-truncated disc, which is characterised by an outer scalelength which is larger than the inner one. The bulge and disc contribute 9 and 78 per cent of the galaxy luminosity, respectively. {\sc gasp2d} does not allow to fit separately the PA for the inner and outer regions of the disc. The best-fitting value of the disc PA ($\rm PA_{\rm disc}=113\fdg0\pm 0\fdg1$) is consistent within errors with the value $\rm PA_{\rm in}=114\fdg0 \pm1\fdg2$ we measured from the isophotal analysis. This is due to the fact that $\rm PA_{\rm disc}$ is driven by the surface brightness distribution of the inner portion of the galaxy, since the image pixels are weighted according to their $S/N$ ratio and {\sc gasp2d} does not allow to fit different values for PA inside and outside the break radius in the case of a double-exponential disc. Although the bar never dominates the galaxy surface brightness, it remarkably contributes 13 per cent of the galaxy luminosity. 

\begin{table}
\centering
\caption{Bulge, disc, and bar structural parameters from the photometric decomposition of NGC~4264.}
\begin{tabular}{cc}
\hline
\multicolumn{2}{ c }{Bulge}\\
\hline
 $\mu_{\rm e}$ & $18.23\pm 0.04$ mag arcsec$^{-2}$\\
 $r_{\rm e}$ & $1.53\pm 0.03$ arcsec\\
 $n$ & $1.38\pm 0.03$\\
 $q_{\rm bulge}$ & $0.77 \pm 0.01$\\
 PA$_{\rm bulge}$ & $96\fdg7 \pm 0\fdg9$\\
 $L_{\rm bulge}/L_{\rm T}$ & 0.09\\
 \hline
\multicolumn{2}{ c }{Disc}\\
\hline
 $\mu_{0}$ & $18.72\pm 0.01$ mag arcsec$^{-2}$\\
 $h_{\rm in}$ & $7.6\pm 0.1$ arcsec\\
 $h_{\rm out}$ & $12.2\pm 0.3$ arcsec\\
 $r_{\rm break}$ & $24.3\pm 0.4$ arcsec \\
 $q_{\rm disc}$ & $0.796\pm 0.002$\\
 PA$_{\rm disc}$ & $113\fdg0\pm 0\fdg1$\\
 $L_{\rm disc}/L_{\rm T}$ & 0.78\\
\hline
\multicolumn{2}{ c }{Bar}\\
\hline
 $\mu_{\rm bar}$ & $19.51\pm 0.01$ mag arcsec$^{-2}$\\
 $a_{\rm bar}$ & $17.31 \pm 0.05$ arcsec\\
 $q_{\rm bar}$ & $0.412\pm 0.001$\\
 PA$_{\rm bar}$ & $56\fdg4\pm 0\fdg1$\\
 $L_{\rm bar}/L_{\rm T}$ & 0.13\\
\hline
\label{tab:structural_parameters}
\end{tabular}
\end{table}

\section{Integral-field spectroscopy}
\label{sec:spectroscopy}

\subsection{Spectra acquisition and reduction}
\label{sec:spectra_reduction}

The spectroscopic observations of NGC~4264 were carried in service mode on 18 and 20 March 2015 (Prog. Id. 094.B-0241(A); P.I.: E.M. Corsini) with MUSE \citep{bacon2010} mounted on the Yepun Unit Telescope 4 of VLT at the Paranal Observatory (Chile) of the European Southern Observatory (ESO).

We configured MUSE in Wide Field Mode and nominal filter. This set up ensured a FOV of $1\times1$ arcmin$^2$ with a $0.2\times0.2$ arcsec$^2$ spatial sampling and wavelength coverage of 4800--9300 \AA\ with a spectral sampling of 1.25 \AA\ pixel$^{-1}$ and a nominal spectral resolution corresponding to ${\rm FWHM}=2.71$ \AA\ at 4800~\AA\ and $2.59$ \AA\  at 9300~\AA .
We split the observations in three observing blocks (OBs) to map the entire galaxy along the photometric major axis for a field coverage of $2.0 \times1.7$ arcmin$^2$. We organised each OB to perform four pointings. The first pointing was on the nucleus of the galaxy and the second one was a sky exposure on a blank sky region at a few arcmins from the galaxy nucleus. The third and fourth pointings were an eastward and westward offset along the galaxy major axis at a distance of 20 arcsec from the galaxy nucleus, respectively. The exposure time of the on-target and on-sky exposures was 780 sec and 300 sec, respectively. In the second and third OB, the pointings were respectively rotated by $90\degr$ and $180\degr$ with respect to the first OB in order to average the spatial signatures of the 24 integral-field units on the FOV. During both nights the seeing reached a mean value of ${\rm FWHM}\sim1$ arcsec. Along with the target and sky observations, day-time (including bias, lamp flatfield, and arc lamp exposures), and twilight calibration exposures (including sky flatfield and spectro-photometric standard exposures) were taken following the standard calibration plan.

We performed the data reduction using the MUSE pipeline \citep{Weilbacher2012} under the {\sc esoreflex} environment \citep{Freudling2013}. The steps included bias and overscan subtraction, lamp flatfielding to correct the pixel-to-pixel response variation of the detectors and illumination edge effects between the detectors, wavelength calibration, determination of the line spread function, sky flatfielding to correct the large-scale illumination variation of the detectors, sky subtraction, flux calibration with correction for atmospheric transmission and differential refraction. 
We combined the twilight flatfield exposures following the same observing pattern of the on-target and on-sky exposure, producing a master twilight datacube to determine the effective spectral resolution and its variation across the FOV. We found an instrumental FWHM$=2.80$~\AA\ ($\sigma_{\rm instr}=69$ km s$^{-1}$) with a negligible variation over the FOV and in the wavelength range between 4800 and 5600 \AA\ which we analysed to measure the stellar kinematics \citep[see also][]{Sarzi2018}.
We estimated the sky contribution by fitting the sky continuum and emission lines on the on-sky exposures. We subtracted the resulting sky model spectrum from each spaxel of the on-target and on-sky exposures. We aligned the sky-subtracted on-target exposures using the common bright sources in the FOV as reference in order to produce a combined datacube of the galaxy.  Even so the resulting sky-subtracted datacube is characterised by a residual sky contamination, which we further cleaned using the Zurich Atmospheric Purge ({\sc zap}) algorithm \citep{Soto2016}. Unfortunately, we were left with residuals from the sky-line subtraction  in the wavelength range centred on the Ca~II$\lambda\lambda8498,8542,8662$ absorption-line triplet.

\subsection{Stellar kinematics and circular velocity}
\label{sec:kinematics}

We measured the stellar and ionised-gas kinematics of NGC~4264 from the sky-cleaned datacube of the galaxy using the Penalized Pixel Fitting ({\sc ppxf}, \citealt{Cappellari2004}) and with the Gas and Absorption Line Fitting ({\sc gandalf}, \citealt{Sarzi2006}) algorithms, which we adapted to deal with MUSE datacube.

We spatially binned the datacube spaxels to increase the $S/N$ ratio and ensure a reliable extraction of the relevant kinematic parameters. We adopted the adaptive spatial binning algorithm by \cite{Cappellari2003} based on Voronoi tessellation to obtain a target $S/N = 40$ for each spatial bin, where the signal and noise are obtained in each spaxel using the spectral range between 4800 and 5600 \AA\ by taking the median of the flux in the wavelength range and the square root of the median of the variance given by the pipeline, respectively. We selected this wavelength range to match the passband of the SDSS image used in the application of the TW method (see Sec.~\ref{sec:omega}).
The resulting spectra are characterised by a maximum $S/N \sim 80$ in the innermost spaxels corresponding to the galaxy centre and a minimum $S/N \sim 20$ in the outermost spatial bins of the galaxy disc. We rebinned each spectrum along the dispersion direction to a logarithmic scale. 

For each spatial bin, we convolved a linear combination of 229 stellar spectra available in the ELODIE library ($R=10000$, $\sigma_{\rm instr}=13$ km s$^{-1}$, \citealt{Prugniel2001}) with a line-of-sight velocity distribution (LOSVD) modelled as a truncated Gauss-Hermite series \citep{Gerhard1993, vanderMarel1993} by a $\chi^2$ minimization.
We selected the stellar spectra to fully cover the parameter space of the effective temperature ($T_{\rm eff}$ from 3000 to 60000 K), surface gravity ($\log{g}$ from $-0.3$ to +5.9 dex), and metallicity ([Fe/H] from $-3.2$ to +1.4 dex) of the ELODIE library and we broadened them to match the MUSE instrumental resolution. After rebinning the stellar spectra to a logarithmic scale along the dispersion direction, we dereshifted them to rest frame and cropped their wavelength range to match the redshifted frame of the galaxy spectra. Moreover, we added a low-order multiplicative Legendre polynomial (degree = 6) to correct for the different shape of the continuum of the spectra of the galaxy and optimal template due to reddening and large-scale residuals of flat-fielding and sky subtraction. We excluded from the fitting procedure the wavelength ranges with a spurious signal coming from imperfect subtraction of cosmic rays and bright sky emission lines.

\begin{figure*}
\includegraphics[scale=0.7, angle=90]{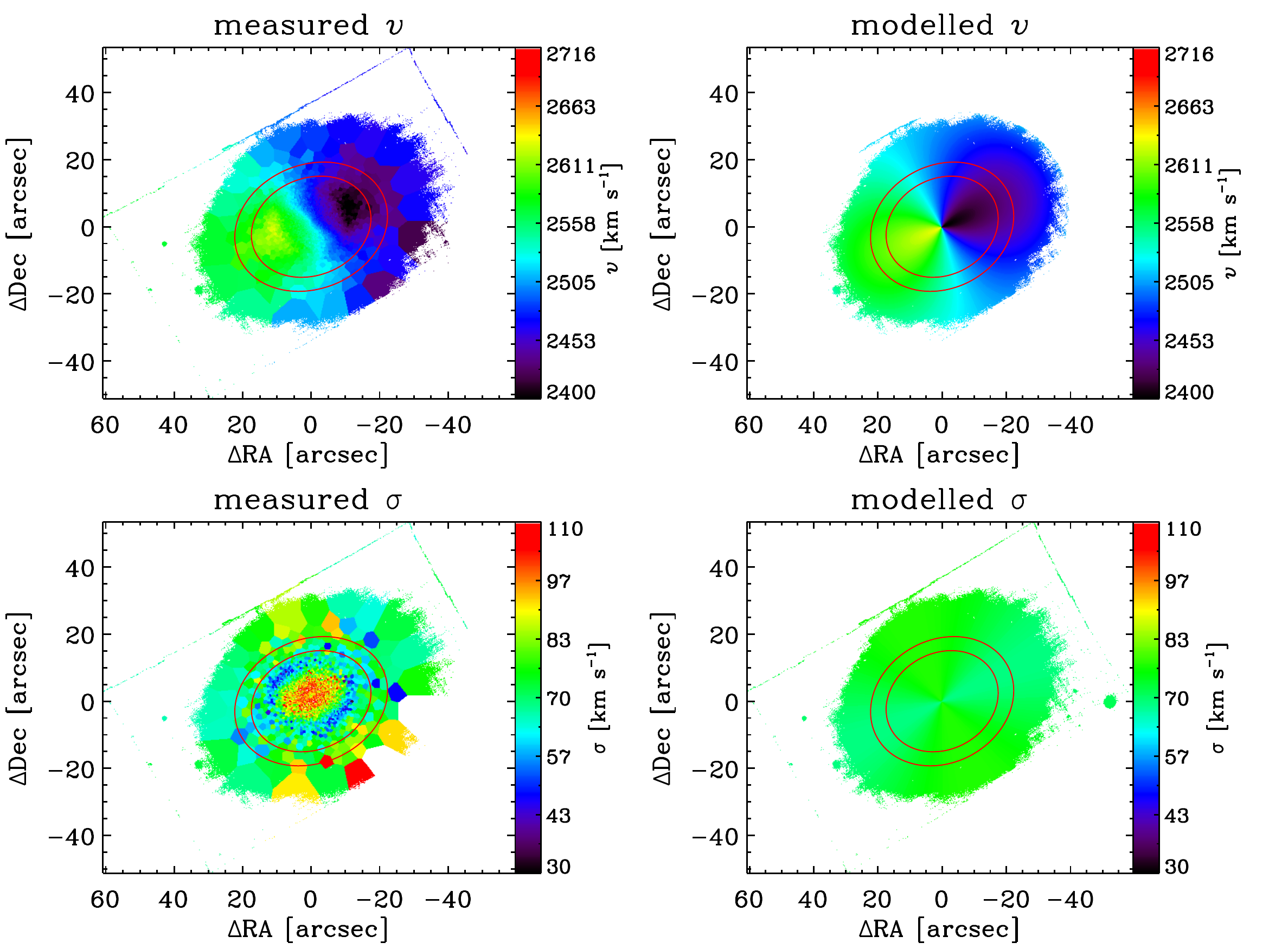}
\caption{Maps of the mean velocity $v$ (upper panels) and velocity dispersion $\sigma$ corrected for $\sigma_{\rm inst}$} (lower panels) of the stars of NGC~4264 derived from the $S/N = 40$ Voronoi binned MUSE data (left-hand panels) and from the asymmetric-drift dynamical model (right-hand panels). The FOV is oriented with North up and East left. The red ellipses bracket the region of the inner disc considered for modelling.
\label{fig:kinematics}
\end{figure*}

By measuring the LOSVD moments in all the available spatial bins in the wavelength range from 4800 to 5600 \AA\ and centred on the Mg\,I$\lambda\lambda5167,5173,5184$ absorption-line triplet, we determined the value of the mean velocity $v$ and velocity dispersion $\sigma$ maps shown in Fig.~\ref{fig:kinematics}. We estimated the errors on LOSVD moments from the the formal errors of the {\sc ppxf} best fit as done in \cite{Corsini2018}. Errors on $v$ and $\sigma$ have ranges between 0.5 and 5 km s$^{-1}$. In addition, we simultaneously fitted with Gaussian functions the ionised-gas emission lines present in the selected wavelength range. The [O\,III]$\lambda\lambda 4959,5007$ and [N\,I]$\lambda\lambda 5198,5200$ emission-line doublets were barely detected in the spectra. Indeed, they have $S/rN \gtrsim 3$, where we estimated the residual noise $rN$ as the standard deviation of the difference between the galaxy and best-fitting stellar spectrum.

In central regions the non-axisymmetric velocity field with an S-shaped zero-velocity isocontour is indicative of the presence of the bar. At larger radii, the regular and axisymmetric velocity field is dictated by the disc component. The velocity dispersion shows a central drop ($\sigma\sim90$ km s$^{-1}$) and a narrow dip ($\sigma\sim45$ km s$^{-1}$) at $R\sim20$ arcsec along the major axis, which corresponds to a ring located just outside the bar region. The central $\sigma$-drop is typical of barred galaxies \citep{Wozniak2006}. We performed a kinemetric analysis of the velocity field using the {\sc kinemetry} algorithm \citep{Krajnovic2006} out to 25 arcsec from the centre finding a good agreement of PA and $\epsilon$ with photometric results. The large bin size and low $S/N$ prevented us from extending our analysis to the outer disc. Our findings are consistent within the errors both with the systemic velocity and the LOS heliocentric velocities obtained in the inner $0.4\times0.7$ arcmin$^2$, given by \cite{Cappellari2011} and \cite{Krajnovic2011}, respectively.

We derived the circular velocity $V_{\rm circ}$ from the LOS stellar velocity and velocity dispersion in the region of inner disc using the asymmetric drift equation \citep{Binney1987}. We selected the spatial bins within an elliptical annulus with semi-major axis between $a=18$ and 23 arcsec and $\epsilon = 0.20$ and followed the prescriptions of \citet{Debattista2002}. We adopted $h_{\rm in}=7.6 \pm 0.1$ arcsec and $i=36\fdg7 \pm 0\fdg7$ for the scalelength of the inner disc from the photometric decomposition and inclination from the isophotal analysis, respectively. We assumed the three components of the velocity dispersion to have exponential radial profiles with the same scalelength but different central values $\sigma_{0,R}$, $\sigma_{0,\theta}$, and $\sigma_{0,z}$, respectively.
This means that the axial ratios of the velocity ellipsoid are $(\sigma_\phi/\sigma_R, \sigma_z/\sigma_R) = (\sigma_{0,\phi}/\sigma_{0,R}, \sigma_{0,z}/\sigma_{0,R})$ and its shape does not change with radius. 
We assumed constant circular velocity, epicyclic approximation ($\sigma_\phi/\sigma_R = 1/\sqrt{2}$), and $\sigma_z/\sigma_R = 0.85  \rm \pm 0.15$ (as typical value for SB0--SB0a galaxies, \citealt{Aguerri2015}). We performed a set of 100 Monte Carlo simulations by varying the values of $h_{\rm in}$, $i$, and $\sigma_{0,z}/\sigma_{0,R}$ within errors and recovering $V_{\rm circ}$ from a Levenberg-Marquardt least-squares fit to the data with the {\sc idl} procedure {\sc mpcurvefit}. We adopted the mean estimate of the circular velocity and corresponding standard deviation as the best-fitting $V_{\rm circ}$ value and associated error, respectively. We found $V_{\rm circ}= 189 \pm 10$ km s$^{-1}$. This estimate is in agreement within the errors with the value $V_{\rm circ}=190.6$ km s$^{-1}$ obtained by \cite{Cappellari2013} by fitting the stellar kinematics of \cite{Krajnovic2011} with a mass-follows-light axisymmetric dynamical model.
The best-fitting kinematic maps of the inner disc are shown in Fig.~\ref{fig:kinematics}.

\section{Characterisation of the bar}
\label{sec:bar}

\subsection{Bar radius}

We obtained the length of the bar semi-major axis $a_{\rm bar}$, which is indicative of the radial extension of the stellar orbits supporting the bar, from the analysis of the SDSS $i$-band image. Since bars do not present sharp edges and often they are associated with other components (like rings or spiral arms) which may affect the bar boundary identification, it is not easy to determine $a_{\rm bar}$ \citep{Aguerri2009}. Several methods have been developed to derive it, but each of them suffers from some limitations \citep[see][for a review]{Corsini2011}. To overcome the problems related to choice of a single measurement method, we derived $a_{\rm bar}$ with three different independent methods, as done for example by \citet{Corsini2003}, \citet{Aguerri2015}, and \citet{Guo2019}.

\begin{figure}
\includegraphics[scale=0.43, angle=90, trim=0 0 30 20]{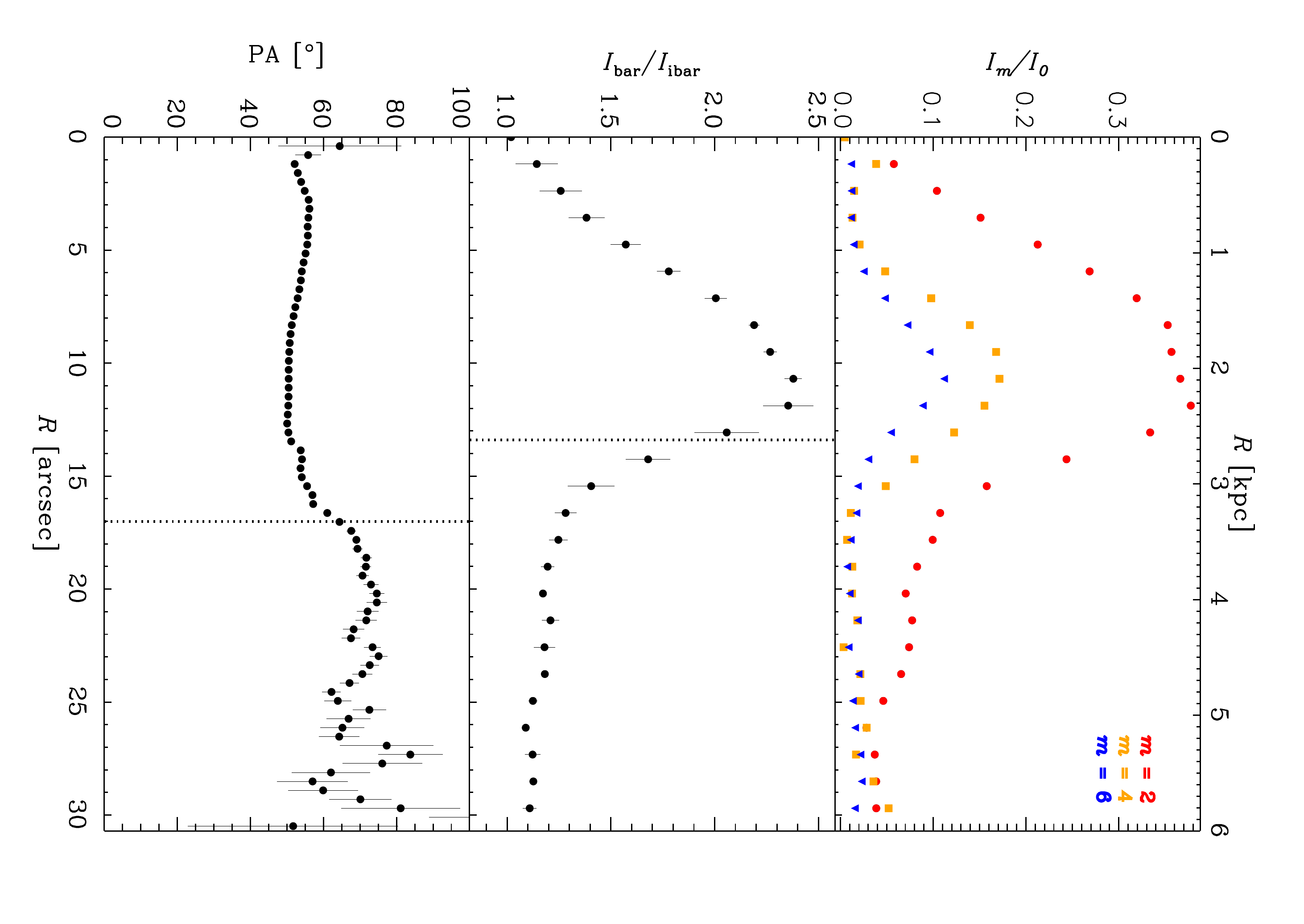}
\caption{Bar radius of NGC~4264 from the analysis of the SDSS $i$-band image. From top to bottom: relative amplitude of the $m=2$ (red circles), 4 (orange squares), and 6 (blue triangles) Fourier component, bar-interbar intensity ratio, and PA of the deprojected best-fitting ellipses. The vertical dotted lines show the value of the bar radius obtained with each method.}
\label{fig:fourier}
\end{figure}

First, we performed a Fourier analysis of the azimuthal luminosity profile of the deprojected SDSS $i$-band image as in \citet{Aguerri2000} (Fig.~\ref{fig:fourier}, top panel). This was obtained through stretching the original image along the minor axis of the galaxy by a factor equal to $\arccos{i}$, where the flux is conserved. The values of PA$_{\rm disc}$ and $i$ are the ones related to the inner part of the disc and recovered in Sec.~\ref{sec:isophotal_analysis}. Through this analysis the bar radius was recovered from the luminosity contrasts between the bar and interbar intensity as a function of radial distance. Our estimate of the bar radius is $a_{\rm bar}=13.4^{+0.2}_{-0.3}$ arcsec (Fig.~\ref{fig:fourier}, middle panel).
A second method to derive the bar radius consists in the analysis of the PA of the deprojected isophotal ellipses \citep{Debattista2002, Aguerri2003}. We obtained the radial profiles of the $\epsilon$ and PA of the deprojected SDSS $i$-band image using {\sc ellipse} and considering a fixed value for the centre of the galaxy. We adopted as bar radius the position where the PA changes by a value of $10\degr$ from the PA of the ellipse with the maximum $\epsilon$ value. A difference of $10\degr$ is a reasonable choice because changing this value between $5\degr$ and $15\degr$ results in bar radius estimates compatible within 1$\sigma$ error. The resulting value is $a_{\rm bar}=17 \pm 3$ arcsec (Fig.~\ref{fig:fourier}, bottom panel).
Finally, we obtained a third estimate for the bar radius $a_{\rm bar}=17.31 \pm 0.05$ arcsec with the photometric decomposition described in Sec.~\ref{sec:gasp2d} (Fig.~\ref{fig:gasp2d}, lower left-hand panel).

We adopted the mean value from the three measurements and the largest deviation from the mean as bar radius and corresponding error, respectively. This gives $a_{\rm bar}=15.9 \pm 2.6$ arcsec.

\subsection{Bar strength}

We obtained the strength of the bar $S_{\rm bar}$, which is an estimate of the non-axisymmetric forces produced by the bar potential, from the analysis of the SDSS $i$-band image using three different methods.

The Fourier analysis allowed us to evaluate the bar strength as defined by \citet{Aguerri2000} and we found $S_{\rm bar}=0.31^{+0.06}_{-0.09}$.
A strictly related definition of bar strength is based on the maximum of the ratio between the amplitudes of the $m=0,2$ Fourier components  \citep{Athanassoula2002, Guo2019} and we obtained $S_{\rm bar}=0.35^{+0.02}_{-0.01}$. 
An alternative estimate for $S_{\rm bar}$ is based on the axial ratio of the bar \citep{Abraham2000} and we derived $S_{\rm bar}=0.27\pm0.01$. 

We adopted the mean value from the three measurements and the largest deviation from the mean as bar strength and corresponding error, respectively. This gives $S_{\rm bar}=0.31\pm0.04$. 

\subsection{Bar pattern speed}
\label{sec:omega}

We derived the bar pattern speed $\Omega_{\rm bar}$ with the model-independent TW method which works for a stellar tracer satisfying the continuity equation and gives
\begin{equation}
\langle X\rangle \: \Omega_{\rm bar} \sin i = \langle V\rangle
\label{eq:tw}
\end{equation}
where $i$ is the disc inclination, while $\langle X\rangle$ and $\langle V\rangle$ are the photometric and kinematic integrals, defined as the luminosity-weighted average of position $X$ and LOS velocity $V_{\rm los}$, respectively, measured along directions parallel to the disc major axis. In long-slit \citep[e.g.,][]{Merrifield1995, Corsini2007} and integral-field spectroscopy \citep{Aguerri2015, Guo2019} $\langle X\rangle$ and $\langle V\rangle$ are derived collapsing the spectrum along the wavelength and spatial directions for each slit and pseudoslit, respectively.

NGC~4264 nicely satisfies all the requirements of the TW method \citep[see][for details]{Corsini2011} because it has an intermediate inclination, its bar is elongated at an intermediate PA between the disc major and minor axes and the disc shows no evidence of spiral arms or patchy dust. If the inclination of the galaxy is too low, LOS velocities are very small and there are large errors associated both with velocities and PA of the disc. On the other side, if the inclination is too high, it becomes difficult to clearly identify the bar and locate the pseudoslits. The bar should be at an intermediate angular location with respect to the disc axes, otherwise it presents nearly zero kinematic or photometric integrals. Spiral arms may lead to a wrong determination of the disc PA and moreover their contribution in light may lead to a contamination of the photometric integrals. The presence of dust may cause a no coincidence between the surface brightness of the galaxy and its mass distribution, which may lead to a mismatch measurement between photometry and kinematics.

The TW method to obtain the bar pattern speed was applied as in Eq.~\ref{eq:tw}. Thanks to the integral-field technique, the pseudoslits were defined a posteriori, from the reconstructed image of NGC~4264.
We defined 9 adjacent pseudoslits crossing the bar (Fig.~\ref{fig:gasp2d}). Each slit has a width of 9 pixels (1.8 arcsec) to deal with seeing smearing effects, a half length of 125 pixels (25 arcsec) to cover the extension of the inner disc and a ${\rm PA}=114\fdg0$ corresponding to the PA of the inner disc, to get a physical solution, as described in Sec.~\ref{sec:bar_rate}.

To measure the photometric integrals of NGC~4264 we analysed the MUSE reconstructed image we obtained by summing the MUSE datacube along the spectral direction in the same wavelength range adopted to measure the
stellar kinematics. For each pseudoslit, we calculated
\begin{equation}
   \langle X\rangle = \frac{\sum_{(x,y)} F(x,y) {\rm dist}(x,y)}{\sum_{(x,y)} F(x,y)}
\end{equation}
where $(x,y)$ are the single pixels in each pseudoslit, $F(x,y)$ is the flux measured in each pixel in the collapsed image and dist$(x,y)$ is the distance of each pixel with respect to the line crossing the centre of the pseudoslit.
By measuring the values of $\langle X\rangle$ as a function of the pseudoslit length from 10 to 45 arcsec, we found they do not converge and discovered that the reconstructed image was affected by a residual contribution of surface brightness due to the nearby galaxy NGC~4261. The sky subtraction of the MUSE data was performed using a dedicated sky datacube, but the choice of the corresponding pointings did not actually take into account for the light contamination due to NGC~4261.

Therefore, we decided to estimate the photometric integrals from the SDSS $g$-band image of NGC~4264, which was obtained in a wavelength range ($\lambda_{\rm eff}=4640.42$ \AA, $\Delta\lambda=1766.72$ \AA, \citealt{Gunn1998}) close to that we are interested in and from which we carefully subtracted the surface brightness contribution of NGC~4261 as explained in Sec.~\ref{sec:image_reduction}. We checked the convergence of the SDSS photometric integrals as a function of the pseudoslit length from 10 to 75 arcsec in the $g$-band.
We modelled the PSF of both the SDSS $g$-band (${\rm FWHM}\sim 1.5$ arcsec) and MUSE reconstructed image (${\rm FWHM}\sim 1$ arcsec) by fitting with a circular Moffat function several stars in the FOV. We deconvolved the SDSS $g$-band image with the Richardson-Lucy method \citep{Richardson1972, Lucy1974} by applying the {\sc iraf} task {\sc lucy}. We evaluated the relative increase of the surface brightness in the pixel corresponding to the galaxy centre and adopted a 5 per cent change in surface brightness as stop condition for the number of iterations. Finally, we convolved the deconvolved SDSS image with the PSF of the MUSE reconstructed image and rebinned the resulting SDSS image to the MUSE pixel scale. We extracted the photometric integrals from the convolved and resampled SDSS image in the pseudoslits we defined on the MUSE reconstructed image.
We estimated the errors on $\langle X\rangle$ with a Monte Carlo simulation by generating 100 mock images of the galaxy. To this aim, we processed the convolved and resampled SDSS image using the {\sc iraf} task {\sc boxcar}. Then, we added to each image pixel the photon noise due to the contribution of both the galaxy and sky background and the read-out noise of the detector to mimic the actual image of NGC~4264. We measured the photometric integrals in the mock images and adopted the root mean square of the distribution of measured values as the error for the photometric integral in each pseudoslit.

To obtain the kinematic integrals, we measured the luminosity-weighted LOS velocity $\langle V\rangle$ from the spectra in the wavelength range between 4800 \AA\ and 5600 \AA, after collapsing each pseudoslit along the spatial directions and applying the same method as described for the stellar kinematics in Sec.~\ref{sec:kinematics}. It should be noticed that this is equivalent to using an explicit luminosity weight because the spaxels with higher signal give higher contribution in the collapsed spectrum and consequently in the $V_{\rm LOS}$ determination of each pseudoslit. This corresponds to measuring
\begin{equation}
    \langle V\rangle = \frac{\sum_{(x,y)}V_{\rm LOS} F(x,y)}{\sum_{(x,y)} F(x,y)}
    \label{eq:kinint}
\end{equation}
We adopted the formal errors provided by {\sc ppxf} as errors in the kinematics integrals, following the same prescriptions described in Sec.~\ref{sec:kinematics}.
We checked the convergence of the kinematic integrals by measuring their values as a function of the length along the pseudoslits from 10 to 45 arcsec. The residual background contributed by NGC~4261 does not affect the kinematic integrals which converge in the inner disc region.

\begin{figure}
\includegraphics[scale=0.35, angle=90]{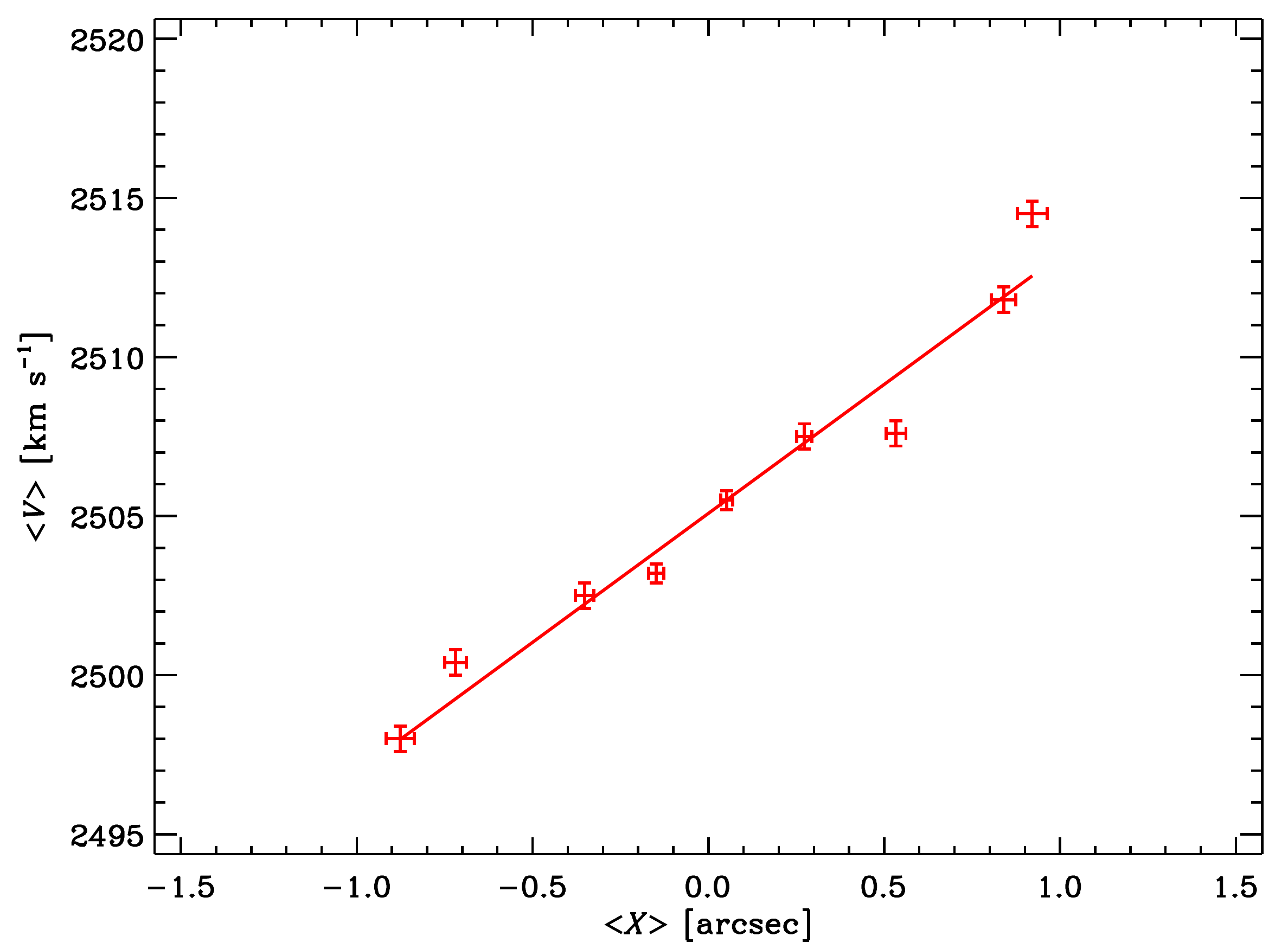}
\caption{Pattern speed of the bar in NGC~4264. The MUSE kinematic integrals $\langle V \rangle$ are plotted as a function of the SDSS photometric integrals $\langle X \rangle$. The best-fitting straight line has a slope $\Omega_{\rm bar} \sin{i} = 8.1 \pm 0.3$ km s$^{-1}$ arcsec$^{-1}$.}
\label{fig:omega_ref}
\end{figure}

We derived $\Omega_{\rm bar}$ of NGC~4264 by fitting a straight line to the photometric and kinematic integrals and their corresponding errors (Fig.~\ref{fig:omega_ref}) using the {\sc fitexy} algorithm \citep{Press1992},  taking into account errors on both $\langle X \rangle$ and $\langle V \rangle$ values. The slope of the best-fitting line is $\Omega_{\rm bar} \sin i$ from which we obtained our reference value of $\Omega_{\rm bar}=13.6 \pm 0.7$ km~s$^{-1}$~arcsec$^{-1}$ (which translates in $71 \pm  4$ km~s$^{-1}$~kpc$^{-1}$).

Although the TW method to derive $\Omega_{\rm bar}$ does not need any modelling, it requires careful measurements to obtain credible values of the photometric and kinematic integrals. We performed a number of tests to scrutinise the different sources of uncertainties on $\Omega_{\rm bar}$ in order to check the reliabily of our reference value. We considered narrower (5 pixels = 1.0 arcsec) and wider (15 pixels = 3.0 arcsec) pseudoslits to halve and double the number of photometric and kinematic integrals to be fitted (test 1). We adopted different PAs for the pseudoslits ($\rm PA_{\rm in}-\sigma_{\rm PA_{\rm in}} = 112\fdg7$, $\rm PA_{\rm in} +\sigma_{\rm PA_{\rm in}} = 115\fdg2$) to account for the uncertainty on the PA of the inner disc (test 2). We measured the kinematic integrals on a larger wavelength range ($4800-5740$ \AA) to verify the kinematic integrals are not affected by the spectral range. This modified spectral interval was selected in order to avoid regions affected by emission or sky residuals and to use a spectral range still similar to the one adopted for the photometric integrals (test 3). We considered only even and odd pseudoslits to deal with fully independent data and minimise the impact of spatial correlations on the photometric and kinematic integrals (test 4). In addition, we adopted the photometric integrals measured on the reconstructed MUSE image to address the amount of light contamination due to NGC~4261 (test 5). The results are presented in Table~\ref{tab:test}.

All the resulting values of $\Omega_{\rm bar}$ given in Table~\ref{tab:test} are consistent within the errors with the adopted reference value, except for the case of test 2. As expected \citep[see][for a discussion]{Corsini2011}, the misalignment between the pseudoslits and disc PA is the main source of uncertainty on $\Omega_{\rm bar}$ and it translates into a systematic error which depends on the bar orientation and disc inclination \citep{Debattista2003}. For NGC~4264, a PA uncertainty of $1\fdg2$ translates into a maximal systematic relative error $\Delta\Omega_{\rm bar}/\Omega_{\rm bar}=0.21$ in agreement with previous findings by \citet{Debattista2004}.

\begin{table}
\centering
\caption{Bar pattern speed and bar rotation rate of NGC 4264.}
\renewcommand{\arraystretch}{1.1}
\begin{tabular}{ ccc }
\hline 
  Parameters  & $\Omega_{\rm bar}$ & {$\cal{R}$} \\
   & [km s$^{-1}$ arcsec$^{-1}$] & \\
\hline
\multicolumn{3}{ c }{Reference value}\\
\hline
  pseudoslit width: 1.8 arcsec  & \\
  disc PA: $114\fdg0$  & $13.6\pm0.7$ & $0.88\pm 0.23$ \\
  spectral range: 4800--5600 \AA   & \\
  \hline
  \multicolumn{3}{ c }{Test 1 - different pseudoslit width}\\
  \hline
  pseudoslit width: 1.0 arcsec & $13.1\pm0.6$ & $0.91\pm 0.24$ \\
  pseudoslit width: 3.0 arcsec  & $12.8\pm0.8$ &  $0.93\pm 0.26$ \\
  \hline
  \multicolumn{3}{ c }{Test 2 - different pseudoslit PA}\\
  \hline
  disc PA: $112\fdg7$ & $15.6\pm0.7$ & $0.77\pm 0.20$\\
  disc PA: $115\fdg2$ & $9.9\pm0.6$ & $1.20\pm 0.34$\\
  \hline
  \multicolumn{3}{ c }{Test 3 - different spectral range}\\
  \hline
  spectral range: 4800--5740 \AA & $13.4\pm0.7$ & $0.87 \pm 0.23$\\
  \hline
  \multicolumn{3}{ c }{Test 4 - even or odd pseudoslits} \\
  \hline
  odd pseudoslits & $12.8\pm0.9$ & $0.93\pm 0.26$ \\
  even pseudoslits & $14.1\pm0.8$ & $0.85\pm 0.23$  \\
  \hline
  \multicolumn{3}{ c }{MUSE photometric integrals}\\
  \hline
  MUSE $\langle X\rangle$ & $13.8\pm0.6$ & $0.86\pm 0.22$ \\
  \hline
  \multicolumn{3}{ c }{Outer disc PA} \\
  \hline
  disc PA: $122\fdg8 $ & $30\pm1$ &  $0.40 \pm 0.10$ \\
  \hline
    \end{tabular}
    \label{tab:test}
\end{table}

Thus, the right identification of the disc PA is crucial for a safe application of the TW. NGC~4264 hosts an upbending disc with a twist of the external isophotes ($\Delta{\rm PA}\sim10^\circ$) moving from the inner to outer regions ($R>27$ arcsec). We also applied the TW method adopting the PA of the outer disc (${\rm PA}_{\rm out}=122\fdg8$) and extracting the kinematic integrals from 45-arcsec long pseudoslits to cover the extension of the outer disc. We found $\Omega_{\rm bar} =30.1 \pm 1.4$ km~s$^{-1}$~arcsec$^{-1}$ ($145.6 \pm 16.3$ km~s$^{-1}$~kpc$^{-1}$). However, this results in an unphysical solution for $\cal{R}$ as discussed in Sec.~\ref{sec:bar_rate}. We were able to recognise this in applying the TW method thanks to the combination of deep SDSS imaging and excellent MUSE integral-field spectroscopy in terms of FOV, spatial sampling, and $S/N$.

\subsection{Bar rotation rate}
\label{sec:bar_rate}

We calculated the length of the corotation radius $R_{\rm cor} = V_{\rm circ}/\Omega_{\rm bar} = 14.0 \pm 0.9$ arcsec of NGC~4264 from the circular velocity and bar pattern speed we estimated from asymmetric drift equation and TW method, respectively.

Finally, we derived the ratio of the length of the corotation radius to the bar semi-major axis which is the bar rotation rate ${\cal{R}}=R_{\rm cor}/a_{\rm bar} = 0.88\pm 0.23$. This value is consistent within errors with the estimates of the bar rotation rates we obtained from the various tests assessing the reliability of $\Omega_{\rm bar}$ (Table~\ref{tab:test}). The PA uncertainty translates into a maximal systematic relative error $\Delta{\cal{R}}/{\cal{R}}=0.38$. We concluded that the bar of NGC~4264 is consistent with being rapidly rotating. 

On the contrary, if we adopt PA$_{\rm out}$ of the outer disc as estimated from the isophotal analysis to derive the bar pattern speed, the corresponding bar rotation rate ${\cal{R}} =  0.40 \pm0.10$ falls into the regime of the bars extending out of the corotation radius which is unphysical \citep{Contopoulos1981}. This is due to the fact that the PA of the outer disc is not representative of the region of the disc where the bar lives but rather of a distortion due to the ongoing interaction with NGC~4261.

\section{Discussion and conclusions}
\label{sec:conclusions}

We measured the broad-band surface photometry and two-dimensional stellar kinematics of NGC~4264, a barred lenticular galaxy at 39.2 Mpc in the region of the Virgo Cluster, to derive the pattern speed of its bar ($\Omega_{\rm bar} = 13.6\pm0.7$ km~s$^{-1}$~arcsec$^{-1}$ or $71 \pm 4$ km~s$^{-1}$~kpc$^{-1}$) and the ratio of the corotation radius to the bar radius (${\cal{R}}= 0.88 \pm 0.23$). We showed that NGC~4264 hosts a strong ($S_{\rm bar}=0.31\pm0.04$) and large bar ($a_{\rm bar} = 15.9\pm2.6$ arcsec or $3.2\pm0.5$ kpc) which nearly extends out to the corotation radius ($R_{\rm cor} = 14.0\pm0.9$ arcsec or $2.8\pm 0.2$ kpc). This means the bar is rotating as fast as it can like nearly all the other bars in lenticulars and spirals measured so far with different methods including TW \citep[see][and references therein]{Elmegreen1996, Rautiainen2008, Corsini2011, Aguerri2015, Guo2019}.

The bar of NGC~4264 has properties typical of bars in lenticular galaxies. The radius and strength are consistent with the median values obtained for SB0 galaxies by \citet{Aguerri2009}, who analysed the SDSS images of a volume-limited sample of about 2100 disk galaxies out to $z=0.04$. They derived the bar semi-major axis as the radius at which the maximum in the bar ellipticity was reached or as the radius at which the PA changes by $5^{\circ}$ with respect to the value corresponding to the maximum ellipticity \citep{Wozniak1995} and estimated the bar strength from the maximum ellipticity \citep{Abraham2000}. The bar rotation rate is consistent within the errors with the mean value calculated by \citet{Aguerri2015} for 17 SB0--SB0/a galaxies with $\Omega_{\rm bar}$ measured with the TW method.

We took advantage of the extended spectral range, fine spatial sampling, large FOV, and superb throughput of the MUSE integral-field spectrograph in combination with wide-field SDSS imaging to deal with the sources of uncertainty in deriving $\Omega_{\rm bar}$ and ${\cal{R}}$ of NGC~4264. We confidently constrained the position and LOS velocity of the galaxy centre, maximized the number and $S/N$ of the spectra extracted from the pseudoslits crossing the bar, carefully derived the orientation and inclination of the galactic disc, accurately measured the bar radius, and recovered the circular velocity by modelling the stellar kinematics. As a result, the values of $\Omega_{\rm bar}$ and ${\cal{R}}$ for the bar of NGC~4264 are amongst the best-constrained ones ever obtained with the TW method. Their statistical relative errors are as small as $\Delta\Omega_{\rm bar}/\Omega_{\rm bar} = 0.06$ and $\Delta{\cal{R}}/{\cal{R}} = 0.26$, respectively. The PA uncertainty translates into a maximal systematic error of 0.21 and 0.38 on $\Omega_{\rm bar}$ and ${\cal{R}}$, respectively.
Although a wrong assessment of the disc PA introduces a systematic error in the application of the TW method, it does not affect all the galaxy measurements in the same way. So when looking at a sample of galaxies, the misalignment between the pseudoslits and disc PA will produce a scatter of the bar pattern speeds and rotation rates rather than a systematic offset with respect to their actual values.

This is a remarkable result not only with respect to early TW measurements based onto long-slit spectroscopy \citep[see][for a  list]{Corsini2011} but also with respect to those recently derived from integral-field spectroscopy \citep{Aguerri2015, Guo2019}. The combined CALIFA and MaNGA sample counts 66 galaxies, of which 10 have ${\cal{R}}<1$ at 95 per cent confidence level. After excluding these ultrafast galaxies, the relative error of $\Omega_{\rm bar}$ measured by averaging among the upper and lower $1\sigma$ statistical errors for the remaining 56 galaxies ranges between $0.03 < \Delta\Omega_{\rm bar}/\Omega_{\rm bar} < 22$ with median value of 0.32. Only 2 galaxies have $\Delta\Omega_{\rm bar}/\Omega_{\rm bar} \lesssim0.06$, i.e. smaller than that of the bar of NGC~4264. As far as ${\cal{R}}$ is concerned, the statistical relative error is $0.19 < \Delta{\cal{R}}/{\cal{R}} < 1.5$ with a median value of 0.43. Only 5 galaxies in the combined CALIFA and MaNGA sample has $\Delta{\cal{R}}/{\cal{R}}\lesssim  0.26$ but none of them have $\Delta\Omega_{\rm bar}/\Omega_{\rm bar}\lesssim0.06$.

The accuracy on bar parameters of NGC~4264 is remarkably close to that of NGC~7079 \citep{Debattista2004}, which has the best-constrained pattern speed ($\Delta\Omega_{\rm bar}/\Omega_{\rm bar}=0.02$) and rotation rate ($\Delta{\cal{R}}/{\cal{R}}=0.21$) ever measured for a bar using the TW method. It represents the first and only use of Fabry-Perot techniques for measuring the two-dimensional stellar kinematics of a barred galaxy. However, handling these kind of data is generally more difficult with respect to the newly developed packages for reducing, analysing, and visualising data from integral-field spectrographs \citep[see][and references therein]{Mediavilla2011}, which are now routinely offered at 4m and 8m-class telescopes and have become a nearly standard tool for the systematic investigation of the structure and dynamics of nearby galaxies \citep[e.g.,][]{Cappellari2011, Sanchez2012, Bundy2015, Sarzi2018}.
  
Our analysis of the MUSE dataset of NGC~4264 represents a pilot study in anticipation of further accurate MUSE measurements of $\Omega_{\rm bar}$ and $\cal{R}$ on a well-defined sample of barred galaxies covering different morphological types and luminosities. This is needed to rigorously test the predictions of numerical simulations about the time evolution of bar radius and pattern speed as a function of gas content, luminous and DM distribution \citep{Weinberg1985, Debattista2000, Athanassoula2003, Athanassoula2013, MartinezValpuesta2017, Algorry2017}. Such a stringent comparison is still a missing piece of information. For example, \citet{Guo2019} did not find any significant correlation between the bar pattern speed and galaxy properties, like the fraction of DM within the galaxy effective radius and the age and metallicity of the stellar populations inside the bar region. But their findings are severely limited by the large statistical relative errors on $\cal{R}$ for the majority of their sample galaxies.

A small misalignment between the direction along which TW integrals are measured and disc major axis may hamper the determination of $\Omega_{\rm bar}$ \citep{Debattista2003}. In this paper, we showed that integral-field spectroscopy
alone can not successfully address this issue but has to be combined with accurate surface photometry to fine tune the extraction of TW integrals. Indeed, we found that the bar of NGC~4264 appears to considerably extend beyond its corotation (${\cal{R}}= 0.40$ $\pm 0.10$) if its pattern speed ($\Omega_{\rm bar}= 30.1\pm1.4$ km~s$^{-1}$~arcsec$^{-1}$) is measured by aligning the pseudoslits with the major axis of the outermost disc (PA$_{\rm out}=122\fdg2 \pm2\fdg4$). This is an unphysical result for a self-consistent bar, which could be due to an incorrect measurement of either $a_{\rm bar}$ or $R_{\rm cor}$ or both. We measured $a_{\rm bar}$ from the SDSS $i$-band image of NGC~4264 with three different methods and they give consistent results within the errors. As a consequence, the problem is related to either $V_{\rm circ}$ or $\Omega_{\rm bar}$ or both. We estimated $V_{\rm circ}$ by correcting for asymmetric drift the LOS velocities and velocity dispersions measured from the MUSE spectra. Our value is in agreement within the errors with the one derived by \citet{Cappellari2013} and based on a mass-follows-light axisymmetric dynamical model of the stellar kinematics measured with the SAURON spectrograph. This means that we can not rely on this alternative measurement of $\Omega_{\rm bar}$. We deduce that the PA and $\epsilon$ of the outermost isophotes of NGC~4264 are not indicative of the actual orientation and inclination of the disc where the bar lives. For this reason, we restricted our analysis of $\Omega_{\rm bar}$ and $\cal{R}$ to the inner disc. 

Finally, the measurement of $\cal{R}$ of the bar of NGC~4264 allowed us to constrain its formation mechanism.
We interpreted the twist of the outer isophotes ($R>27$ arcsec) of NGC~4264, which are characterised by a rotation of the PA ($\Delta {\rm PA} \sim10^\circ$) and no change of $\epsilon$, as suggestive of a warp due to the ongoing interaction with NGC~4261. The limited number of spatial bins of the stellar velocity field at $R>25$ arcsec prevented us from confirming this with a kinemetric analysis.
However, NGC~4264 is at small projected distance from NGC~4261 (3.5 arcmin or 30 kpc) and it is seen through its stellar halo. The two galaxies are probably gravitationally bound with the difference between their systemic velocities ($|\Delta V_{\rm sys,CMB}|=306 \pm 50$ km s$^{-1}$, \citealt{Fixen1996}) consistent with the velocity dispersion of the rich galaxy group they belong to \citep[$\sigma_{\rm group} = 382$ km~s$^{-1}$, ][]{Kourkchi2017}.

The surface-brightness radial profile of the disc of NGC~4264 is upbending in the outer regions ($R>24$ arcsec). The lack of an observed difference between upbending surface-brightness radial profiles in barred and unbarred galaxies \citep{Borlaff2014, ElicheMoral2015} and the fact they are less common in barred galaxies, suggests that they are not likely to be formed by the action of the bar \citep{Debattista2017}. These features are usually explained as the end result of mergers and interactions, which drive outwards migration or direct accretion of part of the stars and dynamically heat the outer region of the disc. An increase of the tangential-to-radial velocity dispersion ratio is expected \citep[see][for a discussion]{Debattista2017}. Unfortunately, we were not able to constrain the shape of the velocity ellipsoid in the disc region since the measured velocity dispersion was close to the instrumental velocity dispersion provided by our MUSE instrumental setup. Moreover, we inspected the SDSS images as well as the residual image of the surface brightness model of NGC~4264 (Fig.~\ref{fig:gasp2d}, upper right-hand panel) without finding any clear-cut evidence of tidal tails. Such an undisturbed morphology of NGC~4264, the fact that NGC~4261 is much more massive than NGC~4264 ($L_{\rm N4261}/L_{\rm N4264}=6$) and their closeness suggest that the interaction between the two galaxies is weak. 

\cite{MartinezValpuesta2017} investigated with N-body numerical simulations the differences between bars resulting from disc instabilities induced by tidal interactions or self-generated internal processes. In agreement with previous findings \citep{Noguchi1987, Salo1991, Miwa1998, Lokas2014}, they found that bars formed through tidal interaction were born and stay slow (${\cal{R}}>1.4$) all along their evolution. The bar rotation rate is found to be ${\cal{R}}\simeq1.4$ only at the end of an interaction occurred over a long timescale. Since the bar of NGC~4264 is fast, we conclude that its formation was due to self-generated internal processes and it was not triggered neither by the recent interaction with NGC~4261 nor by a previous interaction with an other galaxy in the region of the Virgo Cluster. 

\section*{Acknowledgements}
We thank the referee for the constructive report that helped us to improve the paper. We are grateful to M. Bureau, P. T. de Zeeuw, and J. Falc\'on-Barroso for their valuable comments. VC and IP acknowledge support from the Fondazione Ing. Aldo Gini. VC thanks the Instituto de Astrof\'isica de Canarias and the Universidad de la Laguna for hospitality during the preparation of this paper. VC, EMC, EDB, LM, IP, and AP are supported by Padua University through grants DOR1715817/17, DOR1885254/18, and BIRD164402/16. JALA and JMA are supported by the Spanish MINECO grants AYA2017-83204-P and AYA2013-43188-P. VPD is supported by STFC Consolidated grant ST/R000786/1. EI acknowledges the EU Horizon 2020 Marie Sklodowska-Curie grant n. 721463 to the SUNDIAL ITN Network. IP is funded by a PhD fellowship of the Fondazione Cassa di Risparmio di Padova e Rovigo. 




\bibliographystyle{mnras}
\bibliography{biblio} 







\bsp	
\label{lastpage}
\end{document}